\documentclass[twocolumn,aps,showpacs,draft]{revtex4}
\usepackage{graphicx}
\usepackage{dcolumn}
\newcommand{\beq}{\begin{equation}} 
\newcommand{\eeq}{\end{equation}} 
\newcommand{\ds}{\displaystyle} 
\newcommand{\beqar}{\begin{eqnarray}} 
\newcommand{\eeqar}{\end{eqnarray}} 
\begin{document} 
%\draft 
 
\title{Analysis of particle production in ultra-relativistic heavy
ion collisions within a two-source statistical model} 

\author{Zhong-Dao Lu}
\affiliation{China Institute of Atomic Energy,  
P. O. Box 275(18), Beijing 102413, China
\vspace*{1ex}} 
\affiliation{China Center of Advance Science and 
Technology (CCAST), Beijing 100080, China
\vspace*{1ex}} 
\author{Amand~Faessler}
\affiliation{Institut f\"ur Theoretische Physik, Universit\"at 
T\"ubingen, Auf der Morgenstelle 14, D-72076 T\"ubingen, Germany
\vspace*{1ex}}
\author{C.~Fuchs}
\affiliation{Institut f\"ur Theoretische Physik, Universit\"at 
T\"ubingen, Auf der Morgenstelle 14, D-72076 T\"ubingen, Germany
\vspace*{1ex}}
\author{E.E.~Zabrodin}
\affiliation{Institut f\"ur Theoretische Physik, Universit\"at 
T\"ubingen, Auf der Morgenstelle 14, D-72076 T\"ubingen, Germany
\vspace*{1ex}}
\affiliation{Institute for Nuclear Physics, Moscow 
State University, RU-119899 Moscow, Russia
\vspace*{1ex}}

\date{\today} 
 
\begin{abstract} 
The experimental data on hadron yields and ratios in central lead-lead 
and gold-gold collisions at 158 AGeV/$c$ (SPS) and $\sqrt{s} = 130$
AGeV (RHIC), respectively, are analysed within a two-source  
statistical model of an ideal hadron gas. A comparison with the 
standard thermal model is given. The two sources, which can reach
the chemical and thermal equilibrium separately and may 
have different temperatures, particle and strangeness densities, and 
other thermodynamic characteristics,
represent the expanding system of colliding heavy ions, where the hot
central fireball is embedded in a larger but cooler fireball. The 
volume of the central source increases with rising bombarding energy.
Results of the two-source model fit to RHIC experimental data at 
midrapidity coincide with the results of the one-source thermal model 
fit, indicating the formation of an extended fireball, which is
three times larger than the corresponding core at SPS.
\end{abstract}
 
\maketitle 

%\widetext
 
\section{Introduction}
\label{intro}

Searching for the quark-gluon plasma (QGP) is one of the major objectives 
in the study of relativistic heavy ion collisions. The principal 
question is whether the strongly interacting nuclear, or rather  
parton, matter reaches the stage of chemical and thermal equilibrium.
Although this idea was put forward by Fermi 50 years ago \cite{Fe50},
up to now there is no unambiguous test to probe the degree of 
equilibration in the system. One of the possible approaches is to 
study the equilibration process within microscopic models 
\cite{GeKa93,Brplb98,Belk98,Brprc99,SHSX99,Brat00}. 
The more traditional way is to fit macroscopic observables, yields and 
transverse spectra of particles, obtained in experiments to the 
statistical model (SM) of a fully equilibrated hadron gas.
The simplicity of the SM has led to a very abundant literature (see
\cite{SBM,CRSS93,BrSt95,Raf91,Bec96,RGSG91,Sol99,YeGo99,BHS99,ClRe99,
LeRa99,Yen97,Lu98,Zh00,BrMrhic} and references therein). 
As the experimental data became more precise, it has been understood
that the ideal SM does not provide an adequate description of all 
hadron multiplicities \cite{CRSS93}. Particularly,
the yields of pions are usually underestimated while the abundances of
strange particles are overpredicted. Besides, the calculated particle
number density and the total energy density at the stage of chemical
freeze-out are found to be too large \cite{YeGo99}. Therefore,
some modifications to the SM have been proposed. These improvements
consider the following effects: (i) excluded volume effects that lead
to a Van der Waals type equation of state (EOS) due to a
non-zero "hard-core" radius of hadrons \cite{RGSG91,YeGo99,BHS99};
(ii) strangeness suppression, that enters into the distribution functions
via the phenomenological factor $\gamma_S < 1$ \cite{Raf91,Bec96};
(iii) chemical non-equilibrium of light quarks \cite{ LeRa99}. It is
worth noting that the assumption of a single expanding source remains
the basic {\it ad hoc\/} hypothesis of these models. The scenario of
several fireballs has not been employed except a particular case
in which all fireballs have the same chemical potential and the same
temperature \cite{ClRe99,SoHe95} but move with different
collective velocities along the beam direction. In this special case
the particle ratios are, however, not affected by superimposing a
collective longitudinal kinetic energy and are, therefore, identical
to the results of the single fireball scheme. Obviously, if the baryon 
density and/or the strangeness density are not the same everywhere in 
the reaction volume, the scenario with several independent sources 
cannot be reduced to the single source scenario. Our investigations of 
the two-source scenario have been inspired by the experimental 
observation of decreasing of antiproton to proton ratio with rising 
rapidity in the center-of-mass system \cite{na49pap} (the same is true 
for $\bar\Lambda / \Lambda$ ratio also) which results to low net 
baryon densities in the midrapidity range of relativistic heavy-ion 
collisions at SPS energies (158 AGeV/$c$) \cite{na52}. 
In the two-source model the total reaction volume is divided into
two regions, the inner source and the outer source. Each source 
is assumed to be in the chemical and thermal equilibrium, respectively,
and allowed to have different temperature, baryon and strangeness
densities, etc.
This two-source scenario is also in accordance with microscopic model 
calculations. Such microscopic cascade calculations show that the 
central zone between the remnants of the colliding nuclei is more 
baryon dilute compared to the baryon-rich zones of the nuclei residues 
\cite{Brprc99,SHSX99}. Thus, global equilibrium is not reached by the 
whole system, but local equilibrium in the central zone and in the 
peripheral region may still occur separately.

The paper is organized as follows. Description of the two-source model
is given in Sect.~\ref{tsm}. Sections~\ref{sps} and \ref{rhic}
present results of the model fit to the SPS and RHIC data,
respectively. Comparison with the predictions of the standard thermal
model is also given. Finally, conclusions are drawn in 
Sect.~\ref{disc}.
 
\section{Two-source model}
\label{tsm}

Firstly, we briefly sketch some of the basic principles of the 
statistical model of an ideal hadron gas. In the framework of the
grand canonical ensemble, the macroscopic 
characteristics of the system are derived via a set of distribution 
functions (we work with the system of units where 
$c = \hbar = k_B = 1$) 
\beqar 
\ds 
f(p,m_i) &=& \left\{ \exp{\left[ \left(\sqrt{p^2 + m_i^2} -
\mu_B B_i - \mu_S S_i\right) /T \right] } \right. \nonumber\\
&& \left. \pm 1 \right\} ^{-1}\ , 
\label{eq1} 
\eeqar 
where $p$ and $m_i$ are the momentum and the mass of the hadron 
species $i$, $T$ is the temperature, $\mu_B$ and $\mu_S$ are the 
baryon chemical potential and strangeness chemical potential, $B_i$ 
and $S_i$ are the baryon charge and strange charge of hadron $i$. The 
sign ``+" in Eq.~(\ref{eq1}) stands for fermions and sign``$-$" stands 
for bosons. The particle number density $n_i$, the energy density 
$\varepsilon_i$ and the pressure $P$ in the system read as
\beqar 
\ds 
n_i &=&\frac{N_i}{V}=\frac{g_i}{(2\pi)^3}\int_0^{\infty}f(p,m_i)
d^3p\ ,\\
\label{eq2}
\varepsilon_i &=& \frac{E_i}{V}=\frac{g_i}{(2\pi)^3}\int_0^{\infty}
\sqrt{p^2+m_i^2}\, f(p,m_i) d^3p\ ,\\
\label{eq3} 
P &=& \sum_i \frac{g_i}{(2\pi)^3}\int_0^{\infty}
\frac{p^2}{3(p^2+m_i^2)^{1/2}} f(p,m_i) d^3p\ ,
\label{eq4}
\eeqar
with $g_i$ being the degeneracy factor of hadron $i$ and
$V$ the volume of the system.
The entropy density $s = S/V$ is determined via
the Gibbs thermodynamic identity
\beq 
\ds 
s = (\varepsilon +P -\mu_B\rho_B -\mu_S\rho_S)\, T^{-1}\ .
\label{eq5} 
\eeq 
Here $\rho_B$ and $\rho_S$ are the baryon density and strangeness
density, respectively.
 
In the present paper we develop a two-source statistical model (TSM)
of a hadron gas. The model divides the whole reaction zone into two
regions: the outer region (source 1 or S1) and the inner region
(source 2 or S2). Since these sources are assumed to be in local 
thermal and chemical equilibrium, they are allowed to possess 
different temperatures, net baryon densities, chemical potentials, etc.
Furthermore, the strangeness density is no longer kept zero everywhere
as in the SM. Only the total number of strangeness is required to be
zero because of strangeness conservation in strong interactions.

The characteristics of a single fireball can be described by means of
four independent parameters, such as volume $V$, total energy $E$,
net baryon density $\rho_B$ and net strangeness density $\rho_S$. These
parameters enable one to determine the fireball temperature $T$,
the strangeness and baryon chemical potentials $\mu_S$ and $\mu_B$.
Since the total strangeness is zero in heavy-ion collisions,
the number of free parameters in the SM of an ideal hadron gas
is reduced to three, namely, $V$, $E$ (or $\varepsilon = E / V$)
and $\rho_B = N_B / V$. In the two-source model the number of free
parameters increases to seven. Although the net strangeness in each of
the sources can be nonzero, they are linked via the the condition of
total strangeness conservation:
\beq 
\ds
N_{S1} + N_{S2}  = 0 \ .
\label{eq6}
\eeq
Our choice of the free parameters in the two sources is as follows:
Besides temperature, volume, and net baryon density of the two 
sources, i.e. $T_1$, $V_1$, $\rho_{B1}$ and $T_2$, $V_2$, $\rho_{B2}$,
the strangeness density in source 1, $\rho_{S1}$, is considered as
a free parameter.

\section{Hadron production at SPS}
\label{sps}

The baryon yield and ratios of hadrons at midrapidity in central Pb+Pb 
collisions at 158 AGeV are listed in Table~\ref{tab1} together with 
the results of the TSM fit. All hadrons including their resonances 
and possible anti-particles with masses less than 2 GeV/$c^2$ are 
included in the fitting procedure.  No additional constraints such as 
strangeness suppression or excluded volume are assumed except the 
feeding-back effect from resonance decay. For the sake of comparison, 
we check our model by fitting the same set of experimental data with 
the single-source model. The results of the fit are also listed in 
Table~\ref{tab1}. As in the TSM case, these results are obtained 
including the feeding-back effect but without further corrections. 
Compared to the ideal SM, the TSM improves the agreement with the 
experimental data. The thermodynamic quantities obtained from the two 
fits to experimental data are shown in Table~\ref{tab2}.
One can find the sizable differences in temperature and volume
between the two sources. With 155 MeV the temperature in source 2
is significantly higher than that of 117 MeV in source 1.
The volume of the hot source is about $0.7\,V_0$, where $V_0$
is the volume of a lead nucleus, while the volume of the cooler source 
is four times larger. This fact indicates that the two-source object 
can be interpreted as a hot, relatively small core surrounded by a 
cooler and larger halo. 
The net baryon density in the halo is found to be 14
times as large as that in the core. Hence, the major
part of baryons is contained in the outer source, while the inner
source contains almost all antibaryons. The total net baryon charge
equals,  however, the initial number of participants.
\begin{table}
\caption{ 
\label{tab1} 
Baryon yield and hadron ratios at midrapidity for central lead-lead 
collisions at SPS energies and predictions of the single-source and 
two-source statistical models of an ideal hadron gas. 
} 
\vspace*{1ex}
\begin{ruledtabular}
\begin{tabular}{llllcc}   
   & Data & SM & TSM & $\frac{\rm TSM-Data}{\rm Data}\%$ & Ref.\\  
\hline 
$N_B$ (net)                 & 372$\pm 10$  & 371.9 & 372.1& 0 &   
\protect\cite{Hua97} \\    
$K^+/K^- $                 & 1.85$\pm$ 0.1 & 1.97  & 1.87 & 4 & 
\protect\cite{Kan98} \\ 
$\overline{p}/p$           & 0.07$\pm$ 0.01 & 0.069 & 0.058 &-17 & 
\protect\cite{Bea97} \\
$\overline{\Xi}/\Xi$       & 0.249$\pm$ 0.019 & 0.231 & 0.247 &-1& 
\protect\cite{Cali99} \\ 
$\overline{\Omega}/\Omega$ & 0.383$\pm$ 0.081 & 0.427 & 0.405 & 6& 
\protect\cite{Cali99} \\ 
$\overline{\Lambda}/\Lambda$ & 0.128$\pm$ 0.012& 0.130 & 0.137 & 7& 
\protect\cite{And98} \\ 
$\eta/\pi^0$               & 0.081$\pm$0.013& 0.133 & 0.108  & 33 &
\protect\cite{Pei98} \\
$K_S^0/\pi^-$              & 0.125$\pm$0.019& 0.121 & 0.120  &-4&
\protect\cite{Roe97} \\
$K_S^0/h^-$                & 0.123$\pm$0.02 & 0.102 & 0.102  &-17&
\protect\cite{And99} \\
$\Lambda/h^-$              & 0.077$\pm$0.011& 0.069 & 0.064  &-17&
\protect\cite{And99} \\
$\Omega/\Xi$               & 0.219$\pm$0.045& 0.131 & 0.104  &-53&
\protect\cite{And99} \\
$\Xi^-/\Lambda$            & 0.110$\pm$0.01 & 0.156 & 0.115  & 4&
\protect\cite{And99} \\
$\chi^2/DOF$                 &                 & 46/9 & 16/5 &  & \\ 
\end{tabular}
\end{ruledtabular}
\end{table} 
 
\begin{table} 
\caption{ 
\label{tab2}
Thermodynamic characteristics of two-source and single-source models. 
} 
\vspace*{1ex}
\begin{ruledtabular}
\begin{tabular}{lccc}      
      & TSM (S1)  & TSM (S2)  &  SM   \\  
\hline 
$T [MeV]$ & 117 & 155 & 158  \\ 
$V [fm^3]$ & 7250(3$V_0$) & 1705(0.7$V_0$) & 4203 \\ 
$\rho_B [fm^{-3}]$ & 0.048 & 0.014 & 0.088 \\ 
$\rho_S [fm^{-3}]$ & 0.0018 & -0.0075  & 0.0 \\ 
$\epsilon [MeV/fm^3]$  &  104 & 314 & 436 \\ 
$P [MeV/fm^3]$  &  14 & 52 & 67 \\ 
$N_B$ & 348 & 55  & 408 \\ 
$N_{\overline{B}}$ & 0 & 30 & 36 \\ 
$N_B - N_{\overline{B}}$ & 348 & 25  & 372 \\ 
$N_S$ & 34 & 42 & 117 \\ 
$N_{\overline{S}}$ & 21 & 55 & 117 \\ 
$N_S - N_{\overline{S}}$ & 13& -13 & 0 \\ 
$\mu_B [MeV]$ & 460 & 45 & 213 \\ 
$\mu_S [MeV]$ & 68  & 17 & 22 \\   
\end{tabular}
\end{ruledtabular}
\end{table}

The strangeness density is negative in S2 and positive in S1. It 
means that the inner source contains more $s$-quarks than 
$\bar{s}$-quarks. This finding is supported by microscopic model 
calculations \cite{Brprc99}. From the microscopic point of view the 
possible explanation of the phenomenon is as follows: Due to the 
strangeness conservation in strong interactions strange and 
anti-strange particles must be produced in pairs. At SPS energies 
strangeness is mainly carried by kaons. Because of the small 
interaction cross section with hadrons, $K^+$ and $K^0$ are leaving 
the central reaction zone easier than strange particles which carry 
$s$ quarks, e.g., $\Lambda$ and $\overline{K}$, thus leading to a 
negative strangeness density in the midrapidity range.

The energy density in S2 is about three times larger than that in
S1. Such a low energy density in the outer source corresponds to
the energy density at thermal freeze-out rather than at chemical
freeze-out \cite{Brprc99}. In other words, the solution for two 
sources can not be reduced to the one-source picture even in the 
case where exclusive midrapidity data have been used.

\begin{table}
\caption{ 
\label{tab3} 
The same as Table \protect\ref{tab1} but for the data set from
Ref.\protect\cite{YeGo99}.
} 
\vspace*{1ex}
\begin{ruledtabular}
\begin{tabular}{llllcc}   
   & Data & SM & TSM & $\frac{\rm TSM-Data}{\rm Data}\%$ & Ref.\\  
\hline 
$N_B$                      & 372$\pm 10$  & 362.9 & 373.9& 0.5 &   
\protect\cite{Hua97} \\    
$h^-$                      & 680$\pm 50$  & 606   & 659  &-3.2 & 
\protect\cite{Afa96} \\ 
$K_s^0$                    &  68$\pm 10$  & 61.6  & 64.4   &-5.5 &  
\protect\cite{Afa96} \\ 
$\phi$                     & 7.6$\pm 1.1$ & 13.4 & 7.89 & 4& 
\protect\cite{Fri97} \\ 
$p-\overline{p}$           & 155$\pm 20$  & 125   &  147 &-5 &
\protect\cite{Afa96} \\
$K^+/K^- $                 & 1.80$\pm$ 0.1 & 1.99  & 1.74 &-3 & 
\protect\cite{Bor97} \\ 
$\overline{p}/p$           & 0.07$\pm$ 0.01 & 0.065 & 0.069 &-1.5 & 
\protect\cite{Bea97} \\
$\overline{\Xi}/\Xi$       & 0.249$\pm$ 0.019 & 0.220 & 0.229 &-8& 
\protect\cite{Cali99} \\ 
$\overline{\Omega}/\Omega$ & 0.383$\pm$ 0.081 & 0.411 & 0.421 &10& 
\protect\cite{Cali99} \\ 
$\chi^2/DOF$                 &                 & 41/6 & 3.1/2 &  & \\ 
\hline
$T [MeV] $                 &                 & 157  & 114 (S1)  & \\ 
                           &                 &      & 155 (S2)  & \\ 
$V [fm^3]$                 &                 & 4160 &17000 (S1) & \\ 
                           &                 &      & 1920 (S2) & \\ 
$\varepsilon [MeV/fm^3]$   &                 &  422 & 60 (S1)   & \\ 
                           &                 &      & 298 (S2)  & \\ 
\end{tabular}
\end{ruledtabular}
\end{table} 
 
For further investigation we applied the two-source model to fit
the set of experimental data used in Ref.~\cite{YeGo99}, which are
a combination of midrapidity and $4\pi$-data. The data 
and the results of the TSM and SM fit are listed in Table~\ref{tab3}. 
Two important facts can be learned from this comparison. First of all,
it is already known that the ideal SM (without strangeness
suppression and excluded volume effects) underpredicts the number of
negatively charged hadrons $h^-$ and overestimates the yield of
$\phi$ mesons. The problem can be cured by increasing the volume
$V$ and decreasing the temperature $T$ of the source \cite{YeGo99}.
However, the antibaryon to baryon ratios become then completely wrong.
The TSM enables one to get both the correct multiplicities of
$h^-$ and $\phi$ and the correct antibaryon/baryon ratios.
Secondly, the temperatures of both sources remain almost unchanged
when we have shifted from the data set of Table~\ref{tab1} to
the data set of Table~\ref{tab3}, and the temperature of the 
central fireball is very close to the temperature of the entire 
system in the single-source model. 

\begin{table}
\caption{ 
\label{tab4} 
The same as Table \protect\ref{tab3} but for the hadron yields and
ratios in the whole rapidity range.
} 
\vspace*{1ex}
\begin{ruledtabular}
\begin{tabular}{llllcc}   
   & Data & SM & TSM & $\frac{\rm TSM-Data}{\rm Data}\%$ & Ref.\\  
\hline 
$N_B$                      & 372$\pm 10$  & 372.7 & 373.5& 0.4 &   
\protect\cite{Hua97} \\    
$h^-$                      & 680$\pm 50$  & 684   & 673.4&-1 & 
\protect\cite{Afa96} \\ 
$K_s^0$                    &  68$\pm 10$  & 72.5  & 68.8   & 1 &  
\protect\cite{Afa96} \\ 
$\phi$                     & 7.6$\pm 1.1$ & 7.5 & 7.6  & 0 & 
\protect\cite{Fri97} \\ 
$p-\overline{p}$           & 155$\pm 20$  & 138   &  143 &-8 &
\protect\cite{Afa96} \\
$\pi^-/\pi^+ $                 & 1.10$\pm$ 0.05 & 1.0  & 1.0 & -10 & 
\protect\cite{Gue98} \\ 
$K^+/K^- $                 & 1.80$\pm$ 0.1 & 1.84  & 1.80 & 0 & 
\protect\cite{Bor97} \\ 
$K_S^0/\pi^-$              & 0.125$\pm$0.019& 0.121 & 0.118  &-5.5&
\protect\cite{Roe97} \\
$\chi^2/DOF$                 &                 & 1.3/5 & 0.6/1 &  & \\ 
\hline
$T [MeV] $                 &                 & 126  & 117 (S1)  & \\ 
                           &                 &      & 155 (S2)& \\ 
$V [fm^3]$                 &                 &17400 &18500 (S1) & \\ 
                           &                 &      & 1310 (S2) & \\ 
$\varepsilon [MeV/fm^3]$   &                 & 90.6 & 63 (S1)   & \\ 
                           &                 &      & 327 (S2)  & \\ 
\end{tabular}
\end{ruledtabular}
\end{table} 
 
We have also fitted the hadron yields and ratios taken in the whole
available rapidity range, see Table~\ref{tab4}.  Again, the 
thermodynamic characteristics in S2 are far from the averaged 
characteristics of the combined S1+S2 system. For instance, the 
entropy per baryon, $S/A \equiv s/\rho_B$, equals 31 in source 1 but 
increases up to 373 in source 2, mainly due to the low net baryon 
density. The average entropy per baryon $(S_1 + S_2)/(A_1 + A_2)$ is 
37 which is very close to the value 37.8 obtained from the 
one-source SM fit. Results of all three fits favour the idea of
the formation of a compact hot baryon-dilute central zone with the 
following averaged characteristics: temperature $T = 157 \pm 2$ MeV, 
volume $V = 0.6 \pm 0.1\ V_0$, and baryon chemical potential 
$\mu_B = 31 \pm 14$ MeV. The temperature of the halo is much lower, 
i.e., $T_{S_1} = 117 \pm 3$ MeV.

\section{Hadron production at RHIC}
\label{rhic}

Experimental data on hadron yields and ratios in the midrapidity 
range in central gold-gold collisions at $\sqrt{s} = 130$ AGeV
became available recently 
\cite{PHOB2,PHOB1,STAR1,STAR2,STAR3,STAR4,BRAH1,PHEN1}.
These data are listed in Table~\ref{tab5} together with the 
predictions of the SM and TSM. Surprisingly, now the two-source model
divides the total volume in two parts with approximately equal
sizes, temperatures, and other thermodynamic characteristics.
The results of the SM fit and TSM fit are almost identical.
It seems that the volume of the central fireball significantly 
increases by the transition from the SPS energies to the RHIC ones,
and that hadrons detected in the midrapidity region are originated
from a single thermalized source. Its volume is more than 5000 fm$^3$,
that is about three times larger than the corresponding core volume
in Pb+Pb collisions at the SPS,
and the temperature reaches 186 MeV. Such a high temperature value
is close to the value $T = 190$ MeV obtained in \cite{STAR3}, while
being 10 MeV higher than the temperature $T = 175$ MeV obtained in
\cite{BrMrhic}. However, if the multiplicity of negatively charged
hadrons, $h^-$, is excluded from the data set, the temperature of
the modelled system drops to 176 MeV in our calculations. This 
important question should be clarified in future studies. We checked 
also that the incorporation of the excluded volume effects by 
assigning the hard-core radius $r = 0.4$ fm to all particles leads 
to an enlargement of the total volume but does not affect the 
temperature of the fireball.

\begin{table}
\caption{ 
\label{tab5} 
Hadron multiplicities and ratios at midrapidity for central gold-gold 
collisions at RHIC energy ($\sqrt{s} = 130$ AGeV) and predictions of 
the single-source and two-source statistical models of an ideal hadron 
gas.  } 
\vspace*{1ex}
\begin{ruledtabular}
\begin{tabular}{llllcc}   
   & Data & SM & TSM & $\frac{\rm TSM-Data}{\rm Data}\%$ & Ref.\\  
\hline 
$N_B$                      & 343$\pm 11$  & 340.6 & 340.6& -7 &   
\protect\cite{PHOB2} \\    
$h^-$                      &2050$\pm 250$  &2238   &2239  & 9  & 
\protect\cite{PHOB2} \\ 
$\overline{p}/p$           & 0.60$\pm$ 0.07 & 0.57 & 0.57 &-4 & 
\protect\cite{PHOB1} \\
$\overline{p}/\pi^-$       & 0.08$\pm$ 0.01 & 0.087 & 0.087 & 8 & 
\protect\cite{STAR1} \\
$K^-/K^+ $                 & 0.87$\pm$ 0.08& 0.75  & 0.75 &-14 & 
\protect\cite{STAR2} \\ 
$K^-/\pi^-$                & 0.149$\pm$0.02& 0.153 & 0.153  & 3 &
\protect\cite{STAR2} \\
$K^{\ast 0}/h^-$           & 0.060$\pm$0.017& 0.036 & 0.036  &-40 &
\protect\cite{STAR3} \\
$\overline{K^{\ast 0}}/h^-$ & 0.058$\pm$0.017& 0.030 & 0.030  &-48 &
\protect\cite{STAR3} \\
$\overline{\Lambda}/\Lambda$ & 0.77$\pm$ 0.07& 0.69 & 0.69 &-11 & 
\protect\cite{STAR3} \\ 
$\overline{\Xi}/\Xi$       & 0.82$\pm$ 0.08 & 0.80 & 0.80 &-2& 
\protect\cite{STAR4} \\ 
$\chi^2/DOF$               &                & 9.6/7 & 9.6/3 &  & \\ 
\hline
$T [MeV] $                 &                 & 185.7& 185.9 (S1)& \\ 
                           &                 &      & 185.5 (S2)& \\ 
$V [fm^3]$                 &                 & 5157 & 2608 (S1) & \\ 
                           &                 &      & 2560 (S2) & \\ 
$\varepsilon [MeV/fm^3]$   &                 & 1297 & 1305 (S1) & \\ 
                           &                 &      & 1282 (S2) & \\ 
$\mu_B [MeV]$              &                 & 52.6 & 53.3 (S1) & \\ 
                           &                 &      & 51.8 (S2) & \\ 
\end{tabular}
\end{ruledtabular}
\end{table} 
 
Another interesting characteristics at the chemical freeze-out are 
the energy per particle and the entropy per baryon. In \cite{ClRe99} 
the criterion $E/N \approx 1$ GeV was introduced for all particles,
baryons and mesons, in a broad range of bombarding energies 
spanning from a few GeV per nucleon up to several hundred GeV per
nucleon in the center-of-mass system. The predictions of the SM and
the TSM for the midrapidity range are as follows: $E/N = 1.1$ GeV,
and $S/A \equiv s/\rho_B =121$. The last value is about 20\%
below the value $s/\rho_B =150$ predicted by the UrQMD calculations,
but for Au+Au collisions at full RHIC energy, $\sqrt{s} = 200$ AGeV
\cite{LVrhic}. It would be interesting to perform these microscopic 
model calculations also at $\sqrt{s} = 130$ AGeV.  
 
\section{Discussion and Conclusions}
\label{disc}

In summary, the analysis of hadron multiplicities in central
lead-lead and gold-gold collisions at 158 AGeV and $\sqrt{s} = 130$
AGeV, respectively, is presented within a two-source statistical model 
of an ideal hadron gas. 
The TSM fit to the experimental data taken at midrapidity at RHIC
practically coincides with the standard single-source fit. This
result supports the idea of a formation of an extended hot fireball
in the central zone of heavy-ion collisions at RHIC energies.
The temperature of the central fireball varies from 176 MeV to 
185 MeV, depending on the incorporation of the multiplicity of 
negatively charged hadrons, $h^-$, in the fitting data, whereas the 
excluded volume effects seem not to affect the fireball temperature.   
Further going studies are necessary to answer this question. 

At SPS it is found that the properties of the system at chemical 
freeze-out can be well understood in terms of two sources, a central 
core and a surrounding halo, both being in local chemical and thermal 
equilibrium. Temperatures as well as baryon charge and strange 
charge of the two sources are different. 
The thermal characteristics of the central fireball, obtained from the
TSM fit to hadron yields and ratios, depend only weakly on the 
considered rapidity interval of the data, i.e., midrapidity or the 
whole rapidity range.

It is worth mentioning that strangeness seems to be in equilibrium
in both sources which is reflected by $\gamma _S \cong 1$
in our calculations. This observation is in line with the fact
that there is no need to introduce a strangeness suppression factor 
into the standard SM if one fits the particle ratios from the
midrapidity range of Pb+Pb collisions at SPS energies \cite{BHS99}. But 
the factor $\gamma_S < 1$ arises if one intends to fit $4\pi$-data 
\cite{Bec96,YeGo99}. A possible explanation for this puzzle is a
non-homogeneous distribution of the strange charge within the
reaction volume. Therefore, the local, not global, equilibrium of
strangeness can be reached separately in the central and in the
outer part of the expanding fireball.

Furthermore, in the standard SM the energy and hadron number densities
are too high to treat the system at chemical freeze-out
as a gas of point-like particles anymore. To resolve this problem the
introduction of a repulsive hard-core potential for hadrons which
leads to a Van der Waals type EOS might be important
\cite{YeGo99,BHS99}. In the case of two thermalized sources neither the
energy density nor the hadron density is so large and the
incorporation of excluded volume effects (at least for the halo)
becomes less important.

Another interesting fact is that the temperature of source 2 at SPS
energies
is about 10 MeV lower than the temperature predicted by most of 
the single-source statistical models. In the SM such a low 
temperature can be obtained under the assumption of chemical
non-equilibrium (overpopulation) of light $u$ and $d$ quarks
\cite{ LeRa99}. In the latter case the system is not in a state with
maximum entropy, i.e., it is probably produced directly from an
exploding quark-gluon plasma (QGP) and does not undergo further
chemical equilibration in the hadronic phase \cite{ LeRa99}.
However, the ratios of heavy strange baryons, such as 
$\Omega / \Xi$ or $\overline{\Omega} / \overline{\Xi}$, are still
underpredicted.
The TSM allows for another interpretation of the data.
It is quite likely that hot nuclear matter, which forms the core,
spends enough time in the hadronic phase to reach a chemically
equilibrated state. In this case the system does not remember
about the past, e.g., the QGP stage (see also \cite{Ri01}).
This circumstance complicates the detection of the plasma.
It would be very interesting to investigate the forthcoming data on 
Au+Au collisions at RHIC energies ($\sqrt{s}=130$ and 200 AGeV) in
the whole rapidity range within the TSM in order to check (i) the 
increase of the volume of the central fireball at the expenses of 
the halo; (ii) equilibration of strangeness in both sources; (iii)
a possible change of the halo temperature.

\begin{acknowledgments}
We are grateful to L. Bravina for fruitful discussions.
Z.D.L. acknowledges the hospitality of the Institute of Theoretical
Physics, T\"ubingen University. The work was supported by the
Deutsche Forschungsgemeinschaft (DFG), Bundesministerium f\"ur
Bildung and Forschung (BMBF) under the contract No. 06T\"U986, and
the National Science Foundation of China under the contracts
No. 19975075 and No. 19775068.
\end{acknowledgments}


\begin{thebibliography}{9}

\bibitem{Fe50} E.~Fermi,
Prog. Theor. Phys.  {\bf 5}, 570 (1950); 
Phys. Rev. {\bf 81}, 683 (1951).

\bibitem{GeKa93} K.~Geiger and J.I.~Kapusta,
Phys. Rev. D {\bf 47}, 4905 (1993).

\bibitem{Brplb98} L.V.~Bravina {\it et al.\/},
Phys. Lett. B {\bf 434}, 379 (1998); J. Phys. G {\bf 25}, 351 (1999).

\bibitem{Belk98} M.~Belkacem {\it et al.\/}, 
Phys. Rev. C {\bf 58}, 1727 (1998).

\bibitem{Brprc99}  L.V.~Bravina {\it et al.\/},
Phys. Rev. C {\bf 60}, 024904 (1999); 
Phys. Rev. C {\bf 62}, 064906 (2000).

\bibitem{SHSX99} J.~Sollfrank, U.~Heinz, H.~Sorge, and N.~Xu,
Phys. Rev. C {\bf 59}, 1637 (1999).

\bibitem{Brat00} E.L.~Bratkovskaya, W.~Cassing, C.~Greiner, 
M.~Effenberger, U.~Mosel, and A.~Sibirtsev,
Nucl. Phys. {\bf A675}, 661 (2000).

\bibitem{SBM} R.~Hagedorn,
Suppl. Nuovo Cim. {\bf 3}, 147 (1965);
R.~Hagedorn and J.~Rafelski, Phys. Lett. B {\bf 97}, 136 (1980).

\bibitem{CRSS93} J.~Cleymans, K.~Redlich, H.~Satz, and E.~Suhonen,
Z. Phys. C {\bf 58}, 347 (1993); Nucl. Phys. {\bf A566}, 391c (1994).

\bibitem{BrSt95} P.~Braun-Munzinger, J.~Stachel, J.P.~Wessels, and
N.~Xu, Phys. Lett. B {\bf 365}, 1 (1996). 

\bibitem{Raf91} J.~Rafelski, Phys. Lett. B {\bf 62}, 333 (1991).

\bibitem{Bec96} F.~Becattini, Z. Phys. C {\bf 69}, 485 (1996);
F.~Becattini and U. Heinz, Z. Phys. C {\bf 76}, 269 (1997);
F.~Becattini, M.~Gazdzicki, and J.~Sollfrank,
Eur. Phys. J. C {\bf 5}, 143 (1998).
 
\bibitem{RGSG91} D.H.~Rischke, M.I.~Gorenstein, H.~St\"ocker, and
W.~Greiner, Z. Phys. C {\bf 51}, 2210 (1997).

\bibitem{Sol99} J.~Sollfrank, Eur. Phys. J. C {\bf 9}, 159 (1999).

\bibitem{YeGo99} G.D.~Yen and M.I.~Gorenstein,
Phys. Rev. C {\bf 59}, 2788 (1999).

\bibitem{BHS99} P.~Braun-Munzinger, I.~Heppe, and J.~Stachel,
Phys. Lett. B {\bf 465}, 15 (1999).

\bibitem{ClRe99}  J.~Cleymans and K.~Redlich,
Phys. Rev. C {\bf 60}, 054908 (1999).

\bibitem{LeRa99} J.~Letessier and J.~Rafelski,
J. Phys. G {\bf 25}, 451 (1999); Phys. Rev. C {\bf 59}, 947 (1999).

\bibitem{Yen97} G.D.~Yen, M.I.~Gorenstein, W.~Greiner, and S.-N.~Yang,
Phys. Rev. C {\bf 56}, 2210 (1997).

\bibitem{Lu98} Z.-D.~Lu {\it et al.\/}, 
High Energy Phys. Nucl. Phys. {\bf 22}, 910 (1998).
 
\bibitem{Zh00} Z.Y.~Zhang, Y.W.~Yu, C.R.~Ching, T.H.~Ho, and Z.-D.~Lu,
Phys. Rev. C {\bf 61}, 065204 (2000).

\bibitem{BrMrhic} P.~Braun-Munzinger, D.~Magestro, K.~Redlich, and 
J.~Stachel, Phys. Lett. B {\bf 518}, 41 (2001).

\bibitem{SoHe95} J.~Sollfrank and U.~Heinz,
in ``Quark-Gluon Plasma", vol.2, edited by R.C.~Hwa
(World Scientific, Singapore, 1995).

\bibitem{na49pap} J.~B\"achler {\it et al.\/}, NA49 Collab.,
Nucl. Phys. {\bf A661}, 45c (1999).

\bibitem{na52} G.~Ambrosini {\it et al.\/}, NA52 Collab.,
Phys. Lett. B {\bf 417}, 201 (1998).

\bibitem{Hua97} I.~Huang, NA49 Collab., Ph.D. thesis, University of
California at Davis (1997) ({\it unpublished\/}).

\bibitem{Kan98} M.~Kaneta {\it et al.\/}, NA44 Collab.,
Nucl Phys. {\bf A638}, 419c (1998).

\bibitem{Bea97} I.G.~Bearden {\it et al.\/},
J. Phys. G {\bf 23}, 1865 (1997).
 
\bibitem{Cali99}   R.~Caliandro {\it et al.\/}, WA97 Collab., 
J. Phys. G {\bf 25}, 171 (1999).
 
\bibitem{And98} E.~Andersen {\it et al.\/}, WA97 Collab., 
Phys. Lett. B {\bf 433}, 209 (1998).
 
\bibitem{Pei98} T.~Peitzmann, WA98 Collab.,
Proceedings of ICHEP98 (Vancouver, Canada, 1998) 1469.

\bibitem{Roe97} D.~R\"ohrich, NA49 Collab.,
Proceedings of HEP97 (Jerusalem, Israel, 1997) 613.

\bibitem{And99} E.~Andersen {\it et al.\/}, WA97 Collab., 
J. Phys. G {\bf 25}, 171 (1999); Phys. Lett. B {\bf 449}, 401 (1999).
 
\bibitem{Afa96} S.V.~Afanasjev {\it et al.\/}, NA49 Collab.,
Nucl. Phys. {\bf A610}, 188c (1996).

\bibitem{Fri97} S.V.~Afanasjev {\it et al.\/}, NA49 Collab.,
Phys. Lett. B {\bf 491}, 59 (2000).

\bibitem{Gue98} J.~G\"unther, 
Ph.D. Thesis, University of Frankfurt (1998) ({\it unpublished\/}).

\bibitem{Bor97} C.~Borman {\it et al.\/}, NA49 Collab.,
J. Phys. G {\bf 23}, 1817 (1997).

\bibitem{PHOB2} B.B. Back {\it et al.\/}, PHOBOS Collab.,
Phys. Rev. Lett. {\bf 87}, 102303 (2001).

\bibitem{PHOB1} B.B. Back {\it et al.\/}, PHOBOS Collab.,
Phys. Rev. Lett. {\bf 87}, 102301 (2001).

\bibitem{STAR1} J.~Harris {\it et al.\/}, STAR Collab.,
Talk at QM'2001 [Nucl. Phys. {\bf A} (in press)].

\bibitem{STAR2} H. Caines {\it et al.\/}, STAR Collab.,
Talk at QM'2001, [Nucl. Phys. {\bf A} (in press)].

\bibitem{STAR3} N.~Xu {\it et al.\/}, STAR Collab.,
Talk at QM'2001, nucl-ex/0104021 [Nucl. Phys. {\bf A} (in press)].

\bibitem{STAR4} H.~Huang {\it et al.\/}, STAR Collab.,
Talk at QM'2001 [Nucl. Phys. {\bf A} (in press)].

\bibitem{BRAH1} I.G.~Bearden {\it et al.\/}, BRAHMS Collab.,
Phys. Rev. Lett. {\bf 87}, 112305 (2001).

\bibitem{PHEN1} K.~Adcox {\it et al.\/}, PHENIX Collab.,
Phys. Rev. Lett. {\bf 86}, 3500 (2001).

\bibitem{LVrhic} L.V.~Bravina {\it et al.\/},
Phys. Rev. C {\bf 63}, 064902 (2001); J. Phys. G {\bf 27}, 421 (2001).

\bibitem{Ri01} D.H. Rischke, 
Talk at QM'2001, nucl-th/0104071 [Nucl. Phys. {\bf A} (in press)].
 
\end{thebibliography}
\end{document}